# 面向艺术图像创作智能生成的知识图谱：构建标注新体系

郝凯欣[1]，朱可文[2]，邓晃煌[3*]，邱懿武[4]，丁诗莹[4]，丁晨阳[4]，邹宁[3]，李泽健[1]

（1. 浙江大学 软件学院，浙江 宁波 315199；2. 浙江大学 艺术与考古学院，浙江 杭州 310028；
3. 浙江大学 计算机科学与技术学院，浙江 杭州 310028）；4. 杭州造物云技术有限公司，浙江 杭州 310028）
* 通讯作者

**摘　要**：本研究旨在为艺术图像数据集标注构建一个统一、系统且可参照的知识框架，以解决标注过程中因缺乏共同标准而导致的定义模糊与结果不一致问题。为实现这一目标，本研究基于艺术图像的构成法则，结合《论视觉知识》提出的"视觉知识结构化理论"——即视觉知识需通过"典型-范畴""层次结构"实现空间形状、动态关系的精准表达，呼应《文化构成》中"文化元素双重属性"理论深度梳理中西方艺术理论，并开创性地融入中国文化视角，构建了一个层级化、系统化的艺术图像知识图谱。该图谱解构了艺术图像的核心视觉语言，并将中国画特有的空间理论与象征体系，同西方艺术理论进行参照与补充。该图谱将定性艺术概念转化为清晰的结构化框架，不仅符合人类"视觉知识优先于言语知识"的认知规律，更能为AI艺术图像生成、跨文化艺术分析提供可解释、可推理的视觉知识基础，确保标注数据的高质量与一致性，为AI 2.0时代的艺术智能研究提供关键支撑。

**关键词**：艺术图像；知识图谱；中国文化视角；绘画语言；结构化框架

## Knowledge Graph for Intelligent Generation of Artistic Image Creation: Constructing a New Annotation Hierarchy

JIA Kaixin[1], ZHU Kewen[2], DENG Huanghuang[3*], QIU Yiwu[4], DING Shiying[4],

DING Chenyang[4], Zou Ning[3], LI Zejian[1]

（*1.College of Software Engineering, Zhejiang University, Ningbo 315199, China; 2. College of Art and Archaeology, Zhejiang University, Hangzhou 310028,China; 3. College of Computer Science and Technology, Zhejiang University, Hangzhou 310028,China; 4 Hangzhou Zaowu Cloud Technology Co., Ltd., Hangzhou 310028,China*）
*\* Corresponding Author*

**Abstract:** Our study aims to establish a unified, systematic, and referable knowledge framework for the annotation of art image datasets, addressing issues of ambiguous definitions and inconsistent results caused by the lack of common standards during the annotation process. To achieve this goal, a hierarchical and systematic art image knowledge graph was constructed. It was developed based on the composition principles of art images, incorporating the "structured theory of visual knowledge" proposed by

Academician Pan Yun-he in On Visual Knowledge—which states that visual knowledge must achieve precise expression of spatial forms and dynamic relationships through "prototype-category" and "hierarchical structure". Through in-depth review of Chinese and Western art theories and pioneering integration of the Chinese cultural perspective, this graph took shape. The core visual language of art images was deconstructed by this knowledge graph. Meanwhile, the unique spatial theory and symbolic system of Chinese painting were compared with and supplemented by Western art theories. This graph converts qualitative artistic concepts into a clear structured framework. It not only conforms to the cognitive law that "visual knowledge takes precedence over verbal knowledge" in humans but also provides an interpretable and inferential visual knowledge foundation for AI art generation and cross-cultural art analysis. It ensures the high quality and consistency of annotated data, thus offering key support for art intelligence research in the AI 2.0 era.

本研究立足于两大时代趋势的交汇点：一方面，人工智能技术正迈向数据与知识共同驱动、尤其重视视觉及多模态智能发展的 2.0 阶段；另一方面，"文化基因"挖掘与文化遗产数字化保护已成为增强文化自信、推动文明传承的重要国家战略需求。然而在艺术创作这一特定领域，当前以深度学习为代表的人工智能技术仍存在根本性局限：尽管深度神经网络在艺术品风格分类、作者识别等任务上取得了显著进展，但它们本质上是基于统计模式匹配的"黑箱"系统。它们能够识别"是什么"（例如，这是一幅巴洛克风格的画作），却无法解释"为什么"。模型无法理解一幅画作之所以呈现出特定风格，是源于其构图、用光、色彩、笔触等一系列视觉语言元素的内在组合逻辑，更无法洞察这些视觉选择背后深植的文化观念与审美范式。这种现象，即是计算艺术领域亟待弥合的"语义鸿沟"。

为应对这一挑战，本研究构建了"绘画语言知识图谱"（Painting Language Knowledge Graph, PL-KG）。该图谱并非一个简单的图像标签系统，而是一个结构化的、可解释的知识本体。该图谱将数百年艺术史沉淀下的创作法则、技法原理与美学观念，解构为一套层级化、可计算的机器可读语言。其核心架构，是一个从宏观的创作目标到微观的视觉元素的精细化知识体系，旨在为计算机赋予理解艺术"语法"的能力。完整图谱的可视化旭日图见本文第 20 页图 1。

本文的核心论点是：PL-KG 通过对"视觉知识"[1]与"文化构成"[2]两大核心理论进行具体的、可操作的实现，为艺术表达的表征和创作提供了一个兼具有效性与创新性的框架。其首要的、根本性的创新，在于它对中西方视觉范式进行了系统性的、非主从式的融合。它不再将中国画论中的"散点透视"、"留白"等概念作为西方线性透视体系的补充或例外，而是将其置于同等重要的地位，从而构建了一套更具普遍性与包容性的全球视觉语法。本报告将系统性论述 PL-KG 的理论根基、结构设计及其在推动文化数字化与艺术图像创作智能生成方面的深远价值。

# 1 理论基础：从文化缺省到视觉知识

本知识图谱构建的哲学与理论基石，源于潘云鹤教授对现代设计与人工智能的深刻反思，将其"文化构成"理论与"视觉知识"框架相结合，为 PL-KG 提供合理性，并赋予深厚的文化内涵与认知科学基础。这一理论融合，使得图谱的设计理念从根本上超越了纯粹的形式主义分析。

## 1.1 对"文化缺省"的批判与文化构成理论

本研究的起点,是对现代设计中一种普遍存在的"文化缺省"(Cultural Default)现象的批判,这一批判在潘云鹤教授对包豪斯设计理念的反思有所体现。包豪斯作为现代主义设计的摇篮,其核心理念在于探索艺术设计的抽象规律,提倡将作品简化为立方体、球体、圆锥体等基本几何形体及其组合。因其简洁、理性及对工业材料的适应性,在20世纪50年代后逐渐成为主流。然而,当这种几何形式主义泛滥成设计的"默认选项"时,其内在缺陷充分显露:单调性、类同感、以及因剥离特定文化语境而"缺乏人情味"。这种现象的根源被归结为"设计中文化的缺省"[2]。

为了应对"文化缺省","文化构成"(Cultural Constitution)理论被提出[2]。该理论的核心是首先识别并珍视"文化元素"(Cultural Elements)。文化元素被定义为一种文化中最具特色、承载其身份认同的关键部件,例如中国建筑独特的屋顶与梁柱结构、中国汉字与书法、戏剧中的脸谱服饰等。它们是文化的"枝叶",是文化独特性的基本体现。

进而,"文化构成"被定义为一种设计方法学,即有意识地、系统地运用这些文化元素进行创新的过程。它超越了简单的几何形式构成,主张设计应当与深厚的文化母体相结合,从而获得审美的丰富度与情感的唤起力。PL-KG正是这一思想在分析领域的延伸与应用,旨在分析既有艺术品中的文化构成。通过将"汉字与书法""戏剧与服饰""建筑"等明确列为文化元素,并将中西方艺术中独特的视觉法则(如"散点透视"与"线性透视")并置,PL-KG将文化特异性从一个模糊的背景因素,提升为分析的核心维度,从而系统性地回应了"文化缺省"的问题。

## 1.2 视觉知识的体系结构:操作化的理论框架

如果说"文化构成"理论为PL-KG注入了人文主义的灵魂,那么"视觉知识"(Visual Knowledge)理论则为其构建了认知科学与人工智能的骨架。该理论指出,人类的视觉记忆与言语记忆不同,它是一种可被旋转、折叠、扫描和类比的知识形式,具有独特的时空属性[1]。传统的符号型知识表达(如规则、语义网络)难以捕捉这种形象思维的精髓。PL-KG的设计,正是对"视觉知识"理论体系的一次全面操作化实现,它将图谱的层级结构与视觉知识的三大核心要素——视觉概念、视觉命题、视觉叙事——进行了精确的映射。

### 1.2.1 视觉概念

视觉概念是视觉知识的基本单元,其内部具有复杂的结构。PL-KG中的每一个知识节点,都被设计为一个结构化的"视觉概念"。

视觉概念由"典型"和"范畴"(Prototype and Category)构成[1]。在PL-KG中,一个具体的知识节点,如"S形构图",即是其中的"典型"。而该节点的详细定义(适用场景:河流/山路/人物曲线动态;核心作用:增强纵深感)及其量化标准(量化说明:S形占画面长度比≥60%)则共同界定了这一典型的"范畴"。通过这种"典型+范畴"的结构,PL-KG成功地将一个原本定性、模糊的艺术术语,转化为了一个边界清晰、可被计算机理解和应用的形式化实体。

视觉概念本身包含子概念,呈现出层次化的组织关系[1]。PL-KG的多级目录体系正是这一思想的直接体现。例如,"S形构图"这一具体视觉概念,隶属于其上层概念"几何式构图",而"几何式构图"又隶属于"构图类型",最终归于顶层维度"构图生成"。这种自上而下的树状结构,既符合人类认知习惯,也构筑了清晰的视觉知识推理路径。

部分视觉概念还应包含由典型动作构成的元素[1]。在PL-KG中具体体现为,对艺术图像动态元素的描述,例如"笔触节律"下的"急缓节律"(急促笔触=激动情绪,舒缓笔触=平静情绪),以及"书写性笔触"中的"颤抖笔触""旋转笔触"等标签。这些标签超越了画作最终呈现的形态的描述,而是捕捉创作过程中所蕴含的动态趋势与情感张力。

### 1.2.2 视觉命题

视觉命题描述了视觉概念的时空关系[1]。相应

地，PL-KG 框架下的标注，可以将艺术图像的抽象美学特点描述为多个视觉概念及其相互关系。

视觉概念的空间关系通过场景结构(Scene Structures)呈现，这些结构描述了物体之间的位置、距离、内外、上下、左右及前后等方位关系[1]。艺术作品的场景结构，可以通过 PL-KG 中构图、空间、色彩等维度的标签组合来描述。以达·芬奇的《最后的晚餐》为例，其场景结构可以被表征为一组视觉概念的集合，包括"一点透视""焦点构图""相对对称""水平线动线"以及由人物分组形成的多个"三角形构图"等。这些标签共同解释了画作稳定、庄重且富有戏剧性聚焦效果的空间布局。

视觉概念的时间关系则通过动态结构(Dynamic Structure)呈现，描述不同对象的演化、位移、行动等变化。在 PL-KG 中体现为描述节奏、韵律、情绪氛围和笔触动态的标签。

### 1.2.3 视觉叙事

视觉叙事由一组有序的视觉命题构成[1]。比如 PL-KG 中关于"长卷叙事性透视"的定义正是为强叙事性艺术图像的生成提供依据。许多古画长卷"移步换景"的观看方式事实上转化为一段连续的视觉叙事，每一画面段落可被分析为一个独立的视觉命题，而这些命题的有序组合则构成完整的叙事链条。

由此可见，PL-KG 并非凭空构建，而是融合了人文设计领域的"文化构成"理论与基于认知科学与人工智能技术的"视觉知识"框架。这一理论统一不仅体现知识图谱的内在合理性，更是使其超越一般技术工具，成为连接人文与科学的认识论桥梁。

## 2 PL-KG：一个结构化机器可读框架

当前，现有艺术知识图谱大多服务于艺术家、艺术学生、普通大众的创作指导，缺乏专门为人工智能设计、机器可读的结构化艺术知识体系。为填补这一空白，本研究系统性地梳理中西方绘画艺术领域的权威论著，从 20 部经典理论著作、两部美术辞典及 30 余篇期刊论文中，总结提炼出包含以下七大维度的 PL-KG：构图生成、形状体态生成、透视与空间生成、光影生成、色彩关系生成、笔触与肌理生成、边缘关系生成。

基于此构建的 PL-KG，是一个从顶层概念到底层细则的系统性解构框架，它将一幅画作从一个整体的感性体验，拆解为七个人工智能可读、可分析、可比较的核心维度。本章将详细阐述这七大维度的内部结构，并通过对两件分别代表东西方艺术巅峰的杰作——达·芬奇的《最后的晚餐》与范宽的《溪山行旅图》——进行案例分析，以展示该框架的强大解释力与跨文化适用性。这种多维度的组合式分析，也为理解和定义"艺术风格"这一复杂概念提供了一条可计算的路径。在 PL-KG 下，艺术风格不再是一个单一的、模糊的标签，而是由艺术家在一系列视觉维度上做出的独特选择所构成的、可被形式化表达的"配置向量"，为人工智能的阅读、学习、创作打下基石。

### 2.1 艺术图像的构图逻辑标签

构图是"画作的脊梁"[12]，它组织视觉元素，引导观众视线，并为作品的主题与情感奠定结构基础。PL-KG 将构图解构为"构成核心目标""构图类型""画面充盈度""视点结构""视觉引导""视觉平衡"以及"节奏与韵律"等多个子维度。"构成核心目标"体现构图的功能性导向，呼应了胡温希等人对构图与情感表达关系的论述[6]；"构图类型"的分类综合了 Sattarov Farxod[7] 有关对称性的研究、Kaidong W[9] 对几何秩序的系统阐述，以及罗伯茨[12] 归纳的多种画面架构；"画面充盈度"与"视点结构"的设立，则同时涵盖西方基于单点透视的"焦点构图"与中国绘画中特有的"散点构图"[11] 与"留白"[3] 美学；而"视觉引导""视觉平衡"与"节奏与韵律"的设置是基于阿恩海姆[10]、罗伯茨[12] 等人有关视线流动、形状与色彩重复等视觉原则的论著，并形成本图谱的层次化词汇。

为说明本分类方法的强大解释力与适用性，在西方美术领域，我们选取了达·芬奇《最后的晚餐》这一西方古典构图法则的集大成之作举例。

在视点与空间方面，作品采用了严格的一点透

视（或称平行透视），所有建筑线条的延长线最终汇聚于耶稣的头部，使其无可辩驳地成为画面的视觉中心[63]。这种焦点构图的设计，通过透视引导凸显了耶稣在叙事中的核心地位[68]。在平衡与结构方面，画面呈现出高度的对称式构图，特别是背景建筑与人物分组（耶稣两侧各两组，每组三人）均体现了严谨的相对对称。耶稣本人构成了一个稳定的正三角形，而十二门徒表现惊恐或诧异情绪的人物姿势是在三角稳定构图限制下的，形成稳定与动态的和谐统一，体现达·芬奇对画面布局的娴熟控制[63]。桌子的横向线条则构成了一条强烈的水平动线，赋予画面宁静而庄严的基调。在主题传递上，整个构图服务于一个明确的叙事性情节引导目标，即捕捉耶稣说出"你们中有一个人要出卖我了"之后，门徒们瞬间的戏剧性反应。

在中国艺术领域，我们以范宽《溪山行旅图》这一展现中式构图哲学的经典作品为例。在视点与画面充盈度方面，作品的"前景、中景、远景被给予1：3：9的递增比例，意味着把最大的画面留给远景的主峰"[62]，这种体现巨碑般山峰的画面布局，造成了"关键性的视觉跳跃"，正是郭熙在《林泉高致》中所言"自山下而仰山巅，谓之高远"三远法构图中高远的典范[29]。在平衡与引导方面，画面通过巨大的主山与前景的小路形成视觉上的大小平衡。巍峨山顶墨色浓重，一道晶亮的瀑布从主峰旁的黑色直泻而下，观众的视线最后被引导至前景蜿蜒道路上一列渺小的商旅[62]，形成一条由上至下的垂直动线，强化了山势的崇高感。

在主题传递方面，与《最后的晚餐》服务于叙事的构图不同，《溪山行旅图》在画面中心放置巨大山体的目的是抒情性情绪强化，即通过人与自然的大小对比，传达出政治性含义与道家"天人合一"思想中对自然的敬畏[64]。上述内容均可由本文提出的艺术知识图谱词汇所描述，形成结构化标注，辅助文生图大模型的训练和推理应用。

## 2.2 艺术图像的形状体态标签

这一维度关注艺术家如何在二维平面上创造出具有三维实体感、丰富质感和主观表现力的形象。它涵盖了"体感塑造"（体积、结构、质感）、"形状处理"（变形、简化、解构）、"形状类型"（几何形状、有机形状）以及"正形与负形"（图底关系）等子维度。其中，"体感塑造"的分类是基于阿恩海姆[10]的视觉感知理论、约翰内斯·伊顿[13]的色彩理论以及艺术解剖学[5,14]建立；"形状类型"的分类基于塞尚的几何概括理念[23]、"如画"美学对自然形态的推崇[24]。"形状处理"的三个策略均有明确的艺术流派支撑：变形与夸张对应于表现主义[16-18]，简化与概括继承自塞尚的现代主义理念[19,20]，解构与重组则源自立体主义的创作方法[21,22,55,56]。"正形与负形"的分类则建立在阿恩海姆的图底关系理论及其在计算视觉领域的验证基础上[10,58]。相关理论支撑本文提出知识图谱的形状体态层次化词汇的梳理与构建。

仍以达·芬奇《最后的晚餐》举例。在结构与体感方面，画中每一位门徒的身体都展现了艺术家对人体解剖学的精深理解，通过由明到暗的明暗过渡赋予人物坚实的体积感。在形状处理方面，达·芬奇通过各不相同的姿态来传达门徒们在听到消息后的反应，这些夸张而富有表现力的手势与身体扭转使得画面极具张力：大雅各双臂张开，面露惊愕；彼得探出身子情绪激动；腓力则将双手指向胸口，表示出无辜与恳切[68]。

以范宽《溪山行旅图》中对山石形态的塑造为例。在体感塑造中的质感表现方面，范宽所创"雨点皴"，"用笔由蹲而斜上急出，参以雨点般的密点攒簇"[66]，这种短促有力，有走向性的笔触可以很好地塑造阴阳的体积感和北方山石坚硬的表面纹理[66]。在自然有机形状的传达上，与山石树林相比，范宽描绘人物与动物的线条更具灵活性与轻松之感，从而表现出对象的生命活力[67]。在正形与负形方面，雄伟坚实、墨色厚重的山体构成了画面的正形，而环绕其间的留白被巧妙地用作负形，暗示着流动的云雾与空濛的大气，无疑为全局的稳实厚重糅入虚实并存的和谐[67]。可见，本知识图谱的词汇可以大部分描述上述鉴赏内容。

## 2.3 艺术图像的透视与空间生成标签

空间表现是绘画的核心议题，PL-KG 在此维度上的最大创新，便是将西方与中国的透视体系并置，并提出了融合应用的可能性，从而为这两种不同的空间哲学提供了统一的分析框架。

西方线性透视体系在文艺复兴时期由阿尔伯蒂等人系统化，它基于几何学与光学原理，旨在通过数学计算在二维平面上创造一个科学、精确的三维幻觉空间。与此相对，中国画的空间体系则更侧重于表现心灵感知中的意境，它不拘泥于固定的视点，而是追求一种"可游可居"的体验式空间。PL-KG 不仅分别定义了这两种体系的核心概念，如西方的一点透视、两点透视和中国的散点透视、三远法，更重要的是，它通过中西方透视融合应用这一前瞻性标签，为分析那些融合了两种体系的现代艺术作品提供了工具。

表 1 系统性地比较了两种透视体系的差异。

**表 1 透视体系对比**

Tab.1 Comparison of Perspective Systems

| 特征 | 西方线性透视 | 中国概念性透视 |
| --- | --- | --- |
| 理论基础 | 欧几里得几何学与光学原理 | 道家哲学与主观观察 |
| 视点 | 固定的单一视点（"视觉锥"） | 移动的多视点（"散点透视"、"移步换景"） |
| 空间目标 | 创造理性的、可测量的、幻觉般的 3D 空间 | 唤起沉浸式的、体验性的、情感化的空间（"意境"） |
| 叙事功能 | 捕捉单一、凝固的时间瞬间（"快照"） | 展开连续的时空叙事（尤其在长卷中） |
| 核心概念 | 消失点、视平线 | 三远法（高远、深远、平远）、留白 |

## 2.4 艺术图像的光影生成标签

光影是绘画中塑造形态、引导叙事、营造氛围的核心语言。PL-KG 将光影的运用解构为"光影功能""光源与光线"以及"明暗系统"三个层次，形成一个从"为何用光"到"光从何来"再到"如何布光"的完整逻辑链条。

在光影功能方面，达·芬奇以光线引导观众视线聚焦画面核心：耶稣坐在中间，他的脸被身后明亮的窗户映照，在塑造人物立体感的同时，也表达了耶稣作为"世界之光"的寓意。从这个时候开始，光影在绘画上兼具了象征寓意和空间营造的双重作用[69]。在明暗系统方面，叛徒犹大的面部大部分隐没在阴影之中，与明亮下的耶稣形成对比，他的身体前倾，隔绝了来自耶稣的光，这种处理是一种强烈的视觉隐喻。

中国绘画不求形似而求神似，体现在光影方面就是并非模拟单一光源，而是通过墨色的浓淡干湿来表现一种整体的、概念性的光感。在光影功能方面，通过浓重的墨色表现山体的向背关系，通过留白表现云雾与水光，形成了虚实相生、明暗交错的韵律感。在明暗系统上，范宽运用了浑厚的墨色层层积染，形成了丰富的明暗层次，即"墨分五色"。山石的阳面用较淡的墨，而阴面和沟壑则用极重的焦墨，这种明暗平衡的处理方式，使得山体显得异常坚实、厚重，具有强烈的体积感。

## 2.5 艺术图像的色彩关系标签

色彩是绘画中最直接诉诸情感的元素。PL-KG 的色彩体系整合了物理属性（色相、明度、纯度）、心理感知、文化象征与艺术调和法则，旨在将复杂的色彩选择过程结构化。

尽管壁画的色彩因损坏而严重褪色，但从修复后的版本和早期摹本中仍可分析其色彩策略。达·芬奇运用了相对沉稳的中纯度色调和中明度调式，以符合晚餐场景的庄重氛围。耶稣身着红色长袍和蓝色罩袍，红色象征牺牲与神圣之爱，蓝色象征神性与真理，这种色彩的象征意涵在基督教图像学中具有特定意义。门徒们的衣着色彩各异，但通过色调的统一与邻近色调和的运用，维持了画面的整体和谐。

范宽《溪山行旅图》体现中国画特有的"墨分五色"，即唐代张彦远《历代名画记》所言"运墨而五色具"[66]。从前景树木的浓墨，到中景山石的重墨，再到远山与云雾的淡墨和留白，形成了一个从低明度调式到高明度调式的完整光谱。这种浓、淡、干、湿、黑对单色的极致运用，直抵其内在的"气韵"与"真理"，体现了道家"唯道集虚"的

哲学思想[65]。

## 2.6 艺术图像的笔触与肌理标签

笔触与肌理展现艺术家独特的气质与表现手法，是其情感与风格最直接的载体[66]。PL-KG 的创新之处在于，它将西方绘画中强调塑造与质感的笔触，同中国画中强调书写性与气韵的笔法，整合进一个统一的分析框架。

达·芬奇为了在墙壁上实现油画般的细腻过渡和光影效果，放弃了传统的湿壁画技法，采用了一种实验性的蛋彩与油画混合材料直接涂于干燥的石壁上。这种对材料适配的探索，旨在实现他著名的"Sfumato"即晕涂法的效果，通过极其柔和的过渡来模糊轮廓线，使人物仿佛融入光与空气之中。这种技法可以被归类为一种追求极致软边缘的块面笔触。

范宽在创作《溪山行旅图》时，非常注重笔法的运用，不同的景物运用不同的笔法，以显现出山水、树木、人物的差异。山体的塑造主要依赖侧锋用笔，并兼用圆点与方点传达沉重与活力；处理树木的笔法灵活多变，长短线交替，点叶法双钩法兼备，保存树叶的灵活性，还用短且直接的线性笔触画出松树，具有很强的书法意味。而画家在山石上的流水处下笔利落，方向多变，起笔细，收笔粗，旨在传达溪水下落时的洒脱与速度感；对于平滩上的水，因其流动较漫，还有回流的模样，所以画家在此处用笔柔美灵动，回环松弛，刚柔并济[67]。这些笔触不仅塑造形体，更是艺术家内在精神力量的直接抒发，体现了笔触功能中的情感表达与风格符号层面。

## 2.7 艺术图像的边缘关系生成标签

边缘处理是绘画中一项极为精微的技艺，它深刻影响着画面的空间感、光感乃至艺术风格。PL-KG 的独创性在于，它将边缘关系作为一个独立的分析维度，并提炼出中西方在边缘处理理念上的根本差异。

该框架的核心创见是提出了两种不同的边缘处理哲学。

西式边缘随光影变。在西方写实传统中，边缘的清晰或模糊主要取决于光照条件。亮部、受光面的边缘通常更清晰、更硬（硬边缘），而暗部、背光面的边缘则更模糊、更柔和（软边缘），以模拟光线散射的物理效果。伦勃朗、萨金特等大师都是运用这种光影逻辑来塑造形体与空间的典范。

中式边缘随意境变。在中国画中，边缘（即线条）本身就是独立的审美对象，其处理方式完全服务于画面整体的"气韵"与"意境"。艺术家可以根据需要，用清晰的勾勒线条来强调山石的结构感和书写性，也可以用渲染和破墨技法模糊轮廓，以营造湿润、空濛的氛围。其逻辑是观念性的、诗性的，而非物理性的。

PL-KG 通过将这两种理念结构化，并增设"中西方融合边缘处理"子项，为分析跨文化艺术作品中更为复杂的边缘语言提供支撑。

# 3 PL-KG

本章将详细地讲述 PL-KG 的七个维度的内容具体是如何联系到当代的画论主流学说。本章论述从构图生成、形状体态生成、透视与空间生成、光影生成、色彩关系生成、笔触与肌理生成、边缘关系生成等方面开展。

## 3.1 构图生成

构图作为"画作的脊梁"[12]，既影响画面视觉结构，也承载艺术家意图与情感。PL-KG 将"构图生成"设为第一部分，通过多层级解构为宏到微的体系，该体系既植根艺术理论，又具备计算实现的科学性与前瞻性。

构图首要功能是承载并传达作品主题与情感，且有艺术理论支撑。胡温希等人[6]指出，构图呈现的空间特征是艺术图像创作首要特征，还将构图选择与艺术家情感表达倾向关联，为主题传递目标提供依据。罗伯茨[12]则强调构图是"画作的脊梁"，通过排列图形元素赋予画面连贯韵律感与坚实架构。

"构图生成"下分"构成核心目标""构图类型""画面充盈度""视点结构""视觉引导""视觉平衡"

"节奏与韵律",共7个二级目录,各二级目录下持续细分至四级或五级目录。

### 3.1.1 构成核心目标

"构成核心目标"下分"主题传递""视觉平衡""动线引导",以说明构图目标的多样性。

本研究将主题传递细分为"叙事性"与"抒情性",划分源于对艺术流派创作动机的洞察[6],如历史画的叙事导向与风景画的抒情导向构图差异显著。该二元划分可在知识图谱中,将艺术风格和创作意图参数化,如为叙事性配明确视觉动线,为抒情性配具节奏感的色彩分布。

本研究在视觉平衡理论的重要创新,是将"动态平衡"与传统"静态平衡"并列。引用马列维奇等现代主义艺术家实践(如倾斜正方形的动态张力[8]),该分类将古典构图原则扩展至兼容现代、抽象与极简主义的动态策略;阿姆海恩[10]关于视觉经验动态性的观点,也支持"动态平衡"设置。这种"动静并重"框架,让构图生成既能描述稳定和谐的古典画面,也能标注富运动感与内在张力的现代视觉形式,大幅拓宽风格适应性。

动线引导是构图核心目标之一。罗伯茨[12]认为,构图可引领观众视线至视觉中心并使其停留;画师常通过构图引导视线向期望方向延伸或停留,如同欣赏者顺着"摄影师"(画师)的"镜头"(动线),观看其营造的"艺术空间"[11]。

### 3.1.2 构图类型

本研究将构图分为对称式、不对称式及多种几何式,该分类体系在艺术理论中有清晰脉络。17世纪普珊在《阿尔卡迪牧人》中用环状构图构建和谐画面[4];Sattarov Farxod 明确区分对称与不对称构图,阐明对称结构的平衡特性[7];Kaidong W 将几何式构图与现代艺术图像"秩序感"关联,列举正三角、圆形等类型[9];罗伯茨则列举并阐释 S 形、十字形等"画面架构",支撑几何构图的重要性[12]。

### 3.1.3 画面充盈度和视点结构

本研究将四类独特的构图类型"满幅构图""留白构图""散点构图""焦点构图"专门分类,前两者以"画面充盈度"为划分标准,后两者以"视点结构"为划分标准。

"留白"根植东方文化,西方亦有共鸣:罗伯茨[12]的构图原则(如避免边界分散注意力)与留白的呼吸感、聚焦功能相通,留白可视为对负空间的主动运用;《中国美术简史》提及马远《寒江独钓》以大幅空白突出意境[3],其更是承载"虚实相生"道家思想、激发想象的美学观念。

传统西方绘画,尤其文艺复兴后,多采用单点透视的焦点构图,以汇聚线条和视觉中心构建统一三维空间幻觉。

与此相对,中国传统山水卷轴画则用散点透视的散点构图,以移动视点容纳多时空景物,带来"漫游式"体验;史晓澜[11]的分析支持其独立分类,且与焦点构图形成感知差异——前者是时空延展,后者是单点凝视。

### 3.1.4 视觉引导

视觉引导决定观者注视路径与信息接收顺序,本研究将其分为"视觉中心营造""视觉动线设计"两大模块并细化手段,分类有充分理论依据。Sattarov Farxod[7]指出可通过大小、光影区分构图中心;Kaidong W[9]具体列举明暗、色彩、线条、透视引导及黄金分割等方法;罗伯茨[12]支撑最丰富,既将"视觉中心"列为构图核心,又阐述创造景深的手法(直线/空气/交叠透视),强调明暗体块的引导作用,还讨论了调整对比度等策略。

这些引导手段可操作性、可重复性、机器可读性强,因多可转为可测量特征,如亮度梯度、色差、主导线条方向、透视收敛点坐标、黄金分割点位置等。这让"视觉引导"从纯美学描述,转变为兼具理论深度与工程实现可能的分类体系。

### 3.1.5 视觉平衡

本研究将"视觉平衡"分为5种类型,分别是"色彩平衡""明暗平衡""大小平衡""动态平衡"和"位置平衡"。

本研究认为,视觉平衡是可通过视觉心理学与

构图法则解释的客观现象，非纯粹主观感受。Sattarov Farxod[7]提出，调控元素大小、明度、饱和度等视觉重量，可客观塑造主题显著性与视觉层次；罗伯茨[12]高度重视"明暗体块"，称其作用超画面其他部位，且精心安排明暗色块能让画作质变，同时他提及的"不对称布局"及视觉吸引力，也为"动态平衡""位置平衡"提供了范例。

### 3.1.6 节奏与韵律

"节奏与韵律"是构图的时间性维度，影响视觉流动感与情感表达。阿恩海姆[10]强调视觉经验动态性，讨论形状重复与变异构成视觉节奏，为"形状节奏"分类提供支持；胡温希等人[6]论述色彩节奏及笔触对情感与节奏的作用；罗伯茨[12]提出"用节奏分割画面"，肯定其基础构图原则地位，并比喻好构图如歌曲有"节奏和韵律"，印证其作为核心维度的合理性。

本研究将"笔触节律"归为"节奏与韵律"下的分类项，与"色彩节奏""形状节奏"并列。传统上笔触精细分析多用于艺术品真伪鉴定或艺术家风格研究，而本研究将其纳入构图生成范畴，为艺术风格及笔触驱动的艺术图像标注提供新切入点；罗伯茨[12]关于笔触与艺术家个性表达关系的论述，也侧面支持这一创新点——笔触不仅是技术痕迹，更是情感与节奏的直接载体。

## 3.2 形状体态生成

形状与体态的生成是绘画造型的核心，它不仅关乎物象的物理真实性，更承载着艺术家的主观表达与风格建构。本知识图谱围绕"体感塑造""形状类型""形状处理"及"正形与负形"四个维度，构建了一个从具象再现到抽象表达的完整理论框架。

### 3.2.1 体感塑造

体感塑造旨在通过二维平面创造出具有三维实体感的视觉形象，其核心在于对体积、结构和质感的精准表达。

首先，体积感的塑造是体感塑造的基础。阿恩海姆[10]指出，形体的塑造依赖"明暗阶梯"的陡度变化，还强调"笔势"能构建形体内在张力、强化体积感；约翰内斯·伊顿[13]补充，冷暖色调可塑造体积，因为冷暖对比本身就蕴含对空间远近的暗示。这三者共同构成知识图谱中"体积感"塑造的核心方法：明暗、色彩与笔触。

其次，结构表现关注物象内在的骨架与构造逻辑。《西方美术大辞典》指出"艺术解剖"通过分析骨骼肌肉研究人体形态，"了解内部的构造，才能重现外部的表象"[5]。何舟[14]认为，"肌肉与关节的解剖分析是造型的基础"。达·芬奇研究人体解剖、米开朗基罗聚焦肩膝等关键结构，鲁本斯描绘肌肉表现动态力量，均印证人体结构表现的重要性。对于建筑与机械等非生物，艺术家则通过强调框架结构塑造质感，如用横竖线增强建筑立体感，用直线与局部厚涂强化机械的几何美感与硬度感[15]。

最后，量感与质感赋予物象可触知特性。视觉质感是材质属性的视觉呈现，艺术家常刻画金属、木材、丝绸等材质传递情感与感官体验，如金属冷峻坚硬、木材温馨粗糙[53,54]。沈丹妮分析荷兰写实画家亨克·哈勒曼特作品，总结质感刻画规律，例如描绘玻璃需处理背景色彩的映透与反光，木材需刻画粗糙纹理和凹凸细节[53]。

### 3.2.2 形状类型

在绘画造型中，万物形态可以被归纳为两大基本类型：几何形状与有机形状。这种二元划分构成了本知识图谱中形状分类的基础。

几何形状是表现物体坚实感的根本，许多艺术家摒除"随机的细节"，追求"物体永恒的形状"[5]。塞尚是这一理念的经典代表，他1904年致画家信中提出"用圆柱体、球体、锥体来对待自然"[23]，主张将自然多样形式简化为基本几何形体，追求艺术"有意味的形式"。这一思想对野兽派、立体主义、表现主义等现代艺术流派影响深远。因此，知识图谱将圆柱体、球体、圆锥体、立方体等基础几何体作为核心子类，建立在现代艺术造型理论基石之上。

与此相对，有机形状源于自然，代表生命、流

动与不规则性。其美学源头可追溯至 18、19 世纪"如画"美学（Picturesque）指导的风景油画。戴小蛮[24]总结"如画"核心特征：描绘自然景物的不规则性，避免人为"秩序"；忠实再现布局、形态、光影多变的自然物象。这与知识图谱中有机形状子项的定义——"模拟自然物象曲线形态，用流畅曲线笔触表现自然形态"（如人体曲线、河流蜿蜒、云朵流动）高度契合。

### 3.2.3 形状处理

形状处理是艺术家进行主观表达和风格创造的关键手段。本知识图谱将其系统化为"变形与夸张""简化与概括""解构与重组"三种核心策略。

"变形与夸张"通过改变物象比例、形态或姿态，强化情感与视觉张力。该手法在维也纳分离画派、德国表现主义等现代流派中尤为经典[16]。例如，席勒常用拉长形体、扭曲肢体承载爱与死亡的深刻体验[17]，蒙克《呐喊》通过极度扭曲人脸，打造难忘的痛苦符号[18]。从科学角度，阿恩海姆[10]在《艺术与视知觉》中指出，形体"拉长或压缩"会偏离常规形象，产生方向性视觉张力；局部放大或缩小是艺术家调节画面视觉重力的手段，"物体愈大，其重力就愈大"，更易吸引观者。将其细分为比例拉长/压扁、局部放大/缩小、扭曲变形等子类，为精确描述艺术风格与情感强度提供了结构化词汇。

"简化与概括"是现代主义绘画重构视觉秩序的重要方法。职荧君将二者分为两种策略[19]：简化是"简化物象表面细节，把握其本质特征"，归纳是"主观概括杂乱现实场景，用平面化手法重构画面元素"。塞尚艺术被罗杰·弗莱视为简化的典范，他摒弃传统细节，专注表现物体内在结构本质[20]。朱利安·豪赫伯格用信息论解释简化：确定图形组织结构的信息量越小，该图形越易被感知[10]。这意味着简化的程度可以通过计算图像信息熵或边缘密度的降低来度量；几何化归纳则可通过多边形拟合等算法实现。本分类法的创新在于，将艺术美学中的"简化"理念与"归纳"策略系统化，为量化分析艺术家抽象过程提供清晰路径。

"解构与重组"在立体主义绘画中体现得最充分。以毕加索的《亚威农少女》为例，艺术家多角度打散人物形体再重组，背后是"在平面以多视点形状表现物体"的创作主张[21,22]。丰子恺将立体主义的本质拆解为形体的解构与再构两过程，前者是"围着物体绕圈以获得连续的面"的空间问题，后者是"用外接或内切重组"的时间问题[55,56]。这一理论为知识图谱中"解构"与"重组"的并列分类提供了坚实的理论依据。将这一分类引入标注体系，可增强数据集捕捉主观与抽象图像特征的能力。

### 3.2.4 正形与负形

正形（Figure）与负形（Ground）的图底关系，是视觉组织的基本原则，深刻影响画面空间感知与主次区分。阿恩海姆的经典研究为这一概念奠定了心理学基础[10]：通常封闭、面积小、亮度或饱和度高的区域容易被视为主体"图"，而环绕它的面积较大、延伸的区域则被视为背景"底"。但艺术家常刻意打破或反转这一常规，例如将主体人物处理得稀疏、背景线条密集，以实现独特现代艺术效果。这对应知识图谱中"图底反转关系"的特殊分类。

图底关系的科学性源于人类知觉机制。人类知觉范围有限，无法同时清晰把握视野中所有事物，视觉聚焦某物时，对周围环境的感知会减弱。因此有经验的画家构图时，会刻意降低画面中心外物象的色彩纯度与对比度，以突出重点[57]。这一原则在计算视觉领域也获验证：翟威（Wei Zhai）等人将图底理论应用于卷积神经网络（CNN），通过让模型学习与人类认知一致的图形、背景空间构型线索，显著提升其在视觉模糊场景下的感知组织能力[58]。该研究从计算层面证明，将图底关系作为可分析特征引入标注体系具有科学性与合理性。

## 3.3 透视与空间生成

在二维平面上构建三维空间幻觉，是绘画艺术的挑战之一。本研究将空间生成的方法论系统化，整合了基于物理光学的空气透视、基于构图层级的空间层次处理、基于几何学的中西方线性透视体系，以及超越常规的特殊空间表现手法，旨在为这些复

杂的技法提供一个全面的标注框架,支撑机器阅读。

### 3.3.1 空气透视

空气透视,又称大气透视,是通过模拟大气对光线和视觉的影响来表现空间深度的方法。早在15世纪,达·芬奇就对此进行了系统性的阐述[27]。他在笔记中明确记录:"通过大气的差异,人们能够分辨不同建筑物的不同距离",并指出"较远的建筑物应画得轮廓模糊一些,更蓝一些"。这一精辟观察直接指出了因距离增加而导致的两个核心视觉变化:轮廓模糊(清晰度变化)和色彩偏蓝(冷暖变化),同时色彩的鲜艳度也会随之降低(纯度变化)。这为 PL-KG 将空气透视的核心要素分解为冷暖变化、清晰度变化和纯度变化三个维度提供了坚实的史学与理论支撑。

空气透视的科学基础源于物理光学和视觉心理学。达·芬奇的观察已揭示了其物理成因,即"空气本身具有物理属性,会随着距离的增加而影响物体的视觉呈现"[27]。现代科学解释了这是由于空气中的微粒对光线的散射作用(瑞利散射使远景偏蓝)以及水汽等对细节的遮蔽。在心理学层面,阿恩海姆[10]的视知觉理论提出,人类视觉具有一种简化和组织的倾向,对于画面中那些模糊、饱和度降低的远方物象,知觉系统会自然地将其判断为处于更远的空间位置。

在人工智能创作中,应用空气透视法则具有至关重要的意义。通过学习色彩的冷暖梯度、轮廓的清晰度以及色彩纯度的渐变,有效量化艺术家如何在二维平面上营造出具有深度和纵深感的三维空间幻觉,从而增强了人工智能对空间感与真实感表现手法的捕捉能力。

### 3.3.2 空间层次

空间层次的构建涉及一对辩证关系:一方面是通过景别处理来营造深度,另一方面是通过平面化处理来强调二维性。本研究对这两种看似矛盾的策略进行了系统性的分类与阐释。

景别处理是通过前景、中景、远景的区分来组织复杂场景、引导视线和建立画面秩序的经典手段。罗伯茨[12]明确建议,前景物体可以刻画得相对清晰具体以引导视线,中景作为视觉中心通常需要最丰富的细节和最高的对比度,而远景则应适当简化、弱化以推远空间。这种空间组织方式不仅在西方绘画中普遍存在,在中国传统绘画中也得到了巧妙运用,例如长卷画《清明上河图》通过近、中、远景的巧妙布局来展现宏大的城市风貌[11]。这种方法基于视觉注意力分配的规律,即视觉系统会主动组织画面元素,形成图形与背景的关系。

与营造深度相对的是平面化处理,即有意弱化或消除空间纵深,强调画面的平面构成感。本研究的创新之处在于,深刻地认识到"平面化"并非单一目的技法,而是服务于不同艺术目标的策略集合。其一,是现代主义的"媒介自觉",如格林伯格所论,现代绘画的本质在于承认并利用绘画媒介的平面性[32],蒙德里安的几何抽象便是典型。其二,是后印象派的"结构探索",如塞尚通过压缩景深、简化形体来寻求画面内在结构的稳定与和谐[31]。其三,是中国民间美术的"符号表达",如靳之林所指出,剪纸等艺术中的平面化并非为了抽象,而是为了更有效地将物象转化为承载特定文化观念的视觉符号[35]。

通过对不同文化背景和艺术目标下的平面化手法进行系统性并置与比较,本研究为跨文化的视觉语言分析提供了更细致、更深刻的分类基础。

### 3.3.3 线性透视

线性透视是基于几何学原理[26],在二维平面上再现三维空间的核心技法。本研究创造性地将西方与中国的透视体系并置,并提出了融合应用的可能性,为标注和人工智能创作提供清晰的指南。

西方透视体系在文艺复兴时期得以系统化。阿尔伯蒂将其定义为一种科学的再现方法[26],而弗朗切斯卡提出的"比例递减"数学原理,为透视空间的计算提供了可操作的框架[28]。达·芬奇[27]不仅研究了标准的一点、两点透视,还敏锐地观察并记录了当视野范围极大时,画面边缘的直线会出现弯曲的现象,这可被视为对鱼眼透视效果的早期描述。

这一体系通过严格的数学计算来确定物体在画面中的大小、位置和形状变化，为西方写实绘画的发展提供了强有力的科学工具。

与此形成鲜明对比的是中国画独特的空间表现体系。它并非基于固定的几何学原理，而是更侧重于表现心灵感知中的空间和意境。郭熙在《林泉高致》中提出的"三远法"（高远、深远、平远），系统总结了中国山水画构图的三种基本空间范式[29]。而"散点透视"则允许画家在同一个画面中自由移动视点，将不同时空的场景有机地组合在一起，形成一种"移动视角"的视觉经验，如《清明上河图》[36]。史晓澜将这种透视与电影的镜头语言进行类比，深刻阐释了其独特的叙事功能[11]。

多个维度都清晰地展示了二者的差异。首先，在理论基础上，西方线性透视严格植根于欧几里得几何学与光学原理，追求科学的精确性[26-28]；而中国画的概念性透视则更多地源于道家哲学、画家的敏锐观察与主观记忆，强调对自然精神的领悟[11,30,36]。其次，这种理论差异直接导致了观者位置与空间目标的不同。西方透视体系设定了一个固定、单点的观看位置，形成所谓的"视觉锥"[27]，其目标是在二维平面上创造一个理性的、可测量的、幻觉般的三维空间。与此相对，中国画的散点透视则采用了移动的、多视点的观看方式，即"移步换景"[36]，旨在唤起一种沉浸式的、体验性的、充满情感的空间感，也就是所谓的"意境"。再次，在叙事功能上，二者也表现出显著区别。西方线性透视倾向于捕捉一个凝固的时间瞬间，如同"快照"一般，将特定场景固定下来。而中国画的透视，尤其在长卷中，则能够在时空中展开连续的叙事[11]，如同展开一幅"卷轴"，带领观者经历一段旅程。

本研究不仅对这两种体系进行了比较，还提出了"中西方透视融合应用"这一前瞻性分类。它主张在分析现代创作时，可以识别出艺术家如何将西方线性透视的"比例准确性"与中国画透视的"叙事、意境感"相结合。例如，对画面中的核心主体采用西方透视以保证其写实性，而对背景则采用散点透视以增加空间的灵活性与意境。这种融合突破了单一透视体系的文化局限，为艺术图像等媒介的表现维度带来了新的可能性。

### 3.3.4 特殊空间表现

在现代艺术中，透视问题已走向了更为普遍的空间问题[5]。除了常规的透视体系，艺术史上还存在多种基于特定文化观念或心理体验的特殊空间表现形式。本研究将其归纳为"中国画留白空间""西方超现实空间"和"民间美术平面化空间"三类，以便对这些非传统空间进行有效标注。

中国画的留白空间，其意义不仅于构图技巧。王炳根的研究指出，中国画家通过有意留下的空白来激发观者的想象，表达"虚实相生"的哲学思想[34]。南宋马远、夏圭的山水画中，大量的留白营造出空灵、深远的意境。这背后利用了阿恩海姆所描述的视觉"闭合"趋势，即观者会主动在心理上补充未完成的形状或意义，从而参与到作品的完成之中[10]。

西方超现实空间则致力于探索潜意识和梦境世界。达利等艺术家常采用极其写实的技法，来描绘不合逻辑、超越现实的场景组合，如《记忆的永恒》中软化的钟表[33]。这种空间的构建依赖于对传统写实技法（如透视、光影）的熟练运用，以此来描绘看似真实实则荒诞的景象，在观众心理上造成一种真实与梦幻交织的错位感，从而挑战人们对现实的常规认知。

民间美术的平面化空间则是一种更为独特的、基于哲学功能的视觉语法。靳之林[35]在其研究中明确指出，民间美术的平面化处理是"宇宙本体的哲学符号直接诠释"。创作者并非在表达个人感受，而是转译宇宙的法则与秩序。例如，陕西剪纸中的"阴阳鱼"图式，通过彻底删除光影、透视等写实元素，将物象转化为承载"阴阳相合"哲学概念的纯粹符号。这种平面化通过删除具有偶然性的视觉细节，使符号获得了高度的认知效率，符合视觉传达的认知经济性原则。

本研究将这种平面化定义为一种哲学符号的载体，突破了传统上将其仅仅视为一种风格技法的认

知局限，为数据集中文化基因的视觉化转译提供了关键的标注路径。

### 3.4 光影生成

光影不仅是再现客观世界的物理现象，更是绘画中塑造形态、引导叙事、营造氛围的核心语言。本研究将光影的运用解构为"功能""光源"与"明暗系统"三个层次，形成一个从"为何用光"到"光从何来"再到"如何布光"的完整逻辑链条，为光影效果的精细化标注提供理论依据。

#### 3.4.1 光影功能

光影在绘画中的功能是多维度的，本知识图谱将其归纳为"塑造体积与结构""引导视线与构建节奏""营造氛围与情绪"三大核心功能。

首先，"塑造体积与结构"是光影最基本的功能。阿恩海姆奠基性地指出："光与暗在一个表面上的渐变，是眼睛得以感知体积和三维性的关键"[10]。这一科学原理揭示了光影造型的本质。郎瑜沁[37]通过艺术史案例进一步验证了这一点：达·芬奇首次将光影从直觉性实践提升为系统理论，其明暗对比法的主要目的正是实现"科学现实主义"和"三维体积感"；而巴洛克画家卡拉瓦乔则通过激进的暗色法（tenebrism），将光影的造型功能推向极致，他"运用光与暗的极端反差"强化物体的块面转折，使人物形体如雕塑般凸显。郭茂来[39]则从技法层面总结，体积感是通过模拟光照在物体上形成的连续明暗过渡（即"三大面五调子"）来实现的。

其次，光影是引导视线、构建画面节奏的有力工具，也是西方古典油画中"传达情感、营造氛围和叙事的重要工具"[37]。其叙事功能的核心机制便是视线引导。卡拉瓦乔在《圣马太蒙召》中，将光线精准地集中于基督指向马太的手指，而让周围人物隐入黑暗，从而"孤立了叙事瞬间并强化其情感冲击力"。这完美诠释了"光区聚焦视线"的操作逻辑。伦勃朗则发展出更复杂的节奏语法，他在《夜巡》中通过让军官的金色绶带、少女的白色衣领与火枪反光形成跳跃的光点，这些"明暗交替的节奏"引导视线在画面中迂回穿行，赋予静态群像以动态体验。阿恩海姆的视知觉理论[10]为此提供了科学解释："高亮度的区域倾向于在知觉上'前进'，而黑暗的区域则倾向于'后退'"，揭示了光影引导视线的生物学基础。

最后，光影是营造氛围与情绪的关键手段。达·芬奇在其笔记中开创性地将光影视为"营造画面氛围和表达情感的关键元素"，并提出了三种调式的雏形：低调绘画通过强烈的对比创造"戏剧性、深沉或神秘的氛围"；高调绘画通过大面积光亮区域传递"明亮、和谐、宁静的效果"；中间调则追求柔和过渡以实现"最大程度的真实感"[27]。郎瑜沁的跨流派研究实证了这一理论：卡拉瓦乔的暗色法营造了强烈的戏剧性，乔治·德·拉·图尔的烛光场景创造了神秘幽暗的氛围，而夏尔丹则用漫射的含蓄光线实现了静谧的诗意表达[37]。郭茂来最终将这种经验上升为"明度调式"理论，为知识图谱中四类氛围的划分提供了现代术语支撑[39]。

#### 3.4.2 光源与光线

对光源与光线的系统化分类，是实现可控光影标注的前提。本研究从方向、种类和质量硬度三个维度对光源进行了划分。

光源的方向（如侧光、顶光、逆光）决定了物体的明暗分布和体积呈现。郎瑜沁的研究表明，达·芬奇对光源方向的系统性探索是其光影研究的革命性贡献之一[37]。艺术史上的经典作品为不同方向光源的功能提供了范例：卡拉瓦乔《圣保罗的归信》中的侧光塑造了极致的戏剧性体积；达·芬奇《最后的晚餐》中的顶光赋予场景神圣感；而《吉内薇拉·班琪》背景中的逆光则强化了空间层次感。可以说，光源方向与光影功能紧密关联：侧光长于塑造体积，顶光善于营造氛围，逆光利于构建空间。

光源的种类被划分为自然光、人造光和主观光。郭茂来从物理性质上区分了直射光（如日光）与透射光（如烛光）[39]，这分别对应了自然光与人造光。郎瑜沁则补充了超越物理规律的主观光类型：伦勃朗晚期自画像中"来源不明的暖黄光"属于"情绪化光线"；卡拉瓦乔的强光直射是"戏剧性光线"；

超现实绘画中违背物理的照明则归为"超现实光源"[37]。这种分类的科学性在于，自然光与人造光遵循可预测的光学定律，而主观光则服务于艺术家的情感表达，如伦勃朗用光来揭示"描绘对象的内心状态"。

光线的质量硬度（柔光/强光）则影响着画面的质感与氛围。达·芬奇最早对比了不同光线质量的审美差异："生硬的阳光"能塑造清晰的结构但会损失细节，而"阴天的柔光"则能呈现微妙的过渡但会弱化体积[27]。郎瑜沁的实证分析揭示了其在艺术实践中的应用：荷兰静物画家采用"集中的光线"（强光）以锐利地刻画质感，而夏尔丹则用"漫射、含蓄的光线"（柔光）使物体泛起天鹅绒般的暖调[37]。郭茂来指出，光线硬度直接影响氛围表达：强光（高对比、窄过渡）增强戏剧性，而柔光（低对比、宽过渡）则更适用于营造静谧的氛围[39]。

### 3.4.3 明暗系统

明暗系统是对画面整体光影关系的结构化描述，本研究将其分为技法层面的"三大面五调子"和风格层面的"调式对比"。

"三大面五调子"是写实绘画中表现体积感的基础教学体系。李宗津明确将"三大面"（亮部、灰部、暗部）与"五调子"（高光、中间调、明暗交界线、反光、投影）定义为素描教学的核心概念[38]。郭茂来进一步阐释了其功能："体积感是通过模拟光照形成的连续明暗过渡来实现的"[39]。这个系统将复杂的连续光影变化，解构为有限的、可识别的几个关键部分，为学习者和标注者提供了一套清晰、可操作的体积塑造分析流程。李宗津同时警示需避免机械化理解，强调这些名词仅是辅助方法，不能替代对真实光影的观察[38]，这也反向印证了该系统在实践中的基础性与必要性。

调式对比则从整体明度分布的角度定义了画面的光影风格。其理论源头可追溯至达·芬奇的调式理论：低调需要"强烈对比与大块阴影"以营造戏剧性氛围，高调应是"大面积光亮与柔和阴影"以传递明亮宁静，而中间调则追求"完美过渡"以获得真实感[27]。郭茂来将其提炼为现代术语"明度调式"，将高调定义为"以高明度为主"，低调为"以低明度为主"，而全调（对应达·芬奇的中间调）则追求明度层次的均衡分布[39]。郎瑜沁的案例分析清晰地验证了不同调式的应用与效果：卡拉瓦乔《埋葬基督》的低调强化了悲剧性，夏尔丹静物画的高调传递了静谧感，而伦勃朗《犹太新娘》的全调则承载了深刻的心理深度[37]。这一分类为量化分析画面的整体氛围和情绪基调提供了宏观的参数。

## 3.5 色彩关系生成

色彩是绘画中最具情感表现力和视觉冲击力的元素。本研究构建的色彩关系标注体系，是一个整合了物理属性、心理感知、文化象征与艺术调和法则的多维度框架，旨在将复杂的色彩选择过程结构化、理论化，以便进行系统性分析。

### 3.5.1 色彩属性运用

色彩的三大基本属性——色相、明度、纯度，是分析和运用色彩关系的核心维度。郭茂来在其著作中对这三大属性的功能作了精辟的定义：明度是色彩的骨架，决定了画面的素描关系，如伦勃朗作品中戏剧性的光影结构；色相是色彩的身份，定义了色彩的基本面貌，如梵高《向日葵》中标志性的铬黄色；纯度则关乎情绪的强度，如蒙克《呐喊》中血色天空所带来的强烈不安感[39]。达·芬奇的笔记早已揭示了色彩属性间的互动关系，他观察到环境色对物体固有色的影响[27]，奠定了色彩关系理论的基础。

这些属性的感知效果在艺术实践中得到了广泛验证，具有科学性。关于冷暖，阿恩海姆[10]证实了暖色（如红、橙）具有前进感，而冷色（如蓝、绿）具有后退感，这为通过冷暖梯度塑造空间纵深感提供了心理学依据。关于明度，高明度色调通常显得轻盈，如莫奈《睡莲》中的浅紫色雾霭，而低明度色调则显得沉重，如戈雅《黑色绘画》系列的阴郁氛围[40]。关于纯度，高纯度色彩能引发强烈的视觉张力，如马蒂斯《舞蹈》中的朱红色，而低纯度色彩则能营造静谧、内敛的氛围，如莫兰迪的灰调静

物画[41]。

本研究将色彩属性进行系统分类，为标注工作提供了一套结构化的调色逻辑分析工具，使色彩选择的分析从感性直觉变为一种可推理的"结构性决策"。

### 3.5.2 色彩心理与象征

《西方美术大辞典》[5]介绍"色彩"时明确指出："从我们的文明起源开始，从古老的中国到西欧，几乎到处都存在着对色彩的约定俗成的象征意义。"色彩不仅是物理现象，更是承载情绪与文化意义的符号系统。歌德在其色彩理论中提出，色彩能够直接激发人的情绪：红色象征激情，如德拉克洛瓦《自由引导人民》中高扬的革命旗帜；蓝色则常隐喻忧郁，如毕加索蓝色时期作品中的孤寂人物[40]。阿恩海姆从格式塔动力学的角度解释了暖色的扩张感与积极情绪的关联[10]，为色彩的情绪表达提供了心理学解释。

更深层次地，色彩的象征意义往往与特定的文化语境紧密相连[40]。例如，在中国传统文化中，红色象征喜庆与权力，如故宫的朱墙；而在西方，代表贵族与神性的往往是紫色，如提香画中圣母的紫袍。然而，也存在跨文化的共通象征，例如金色在佛教壁画与拜占庭艺术中都代表着神圣。明确这些共通性与差异性，对于避免在跨文化艺术分析中产生色彩的误读至关重要。

本研究将色彩的文化象征意义进行系统梳理，分离出"东西方共通象征"（如金、白、黑）与"特色象征"（如东方的黄、紫，西方的绿、蓝），为跨文化背景下的艺术作品标注构建了一套清晰的符号选择指南。

### 3.5.3 色彩调和

色彩调和是组织画面色彩关系，使其达到"秩序、统一与和谐"状态的艺术方法。张超与朱晓君[41]系统地提出了实现色彩调和的四种主要路径：色相主导的调和，如利用互补色对比达到生理上的视觉平衡；明度主导的调和，如利用中明度的稳定性统一画面；纯度主导的调和，如通过降低纯度来缓冲色彩冲突；以及面积对比的调和，即通过调整色彩面积的比例来协调强对比色。

这些调和理论具有很强的科学性与可操作性。例如，互补色调和可以依据歌德色环上相距180°的对比关系进行精确操作，梵高在《夜间咖啡馆》中对红绿对冲的运用即是经典案例[41]。色彩调和理论为分析看似冲突的色彩搭配提供了行之有效的解决方案。此外，王峰的研究还指出了材料对色彩调和的影响[42]，天然材料（如木、石）的固有色彩需要通过对比来强化其美感，而人工材料则可以更自由地调节。

本研究将材料载体的物理属性（如油画的厚重感-水墨的透明性）纳入调和体系，补全了在调和体系中"材料调和"的缺位情况，为后续数据集的标注提供更细致的标签。

### 3.5.4 色彩质感与材料适配

色彩最终需要通过物理材料来呈现，材料的特性深刻地影响甚至决定了色彩的最终视觉效果。王峰在其研究中强调，材料是色彩的物理载体，艺术创作必须充分考虑材料本身的质感与美感[42]。例如，石材或木材的天然肌理与色彩，本身就具有审美价值，创作时应予以保留和强化，如亨利·摩尔雕塑中的粗犷赭石[42]。而对于人工材料，则可以通过对色彩属性的调节来烘托其质感，如马蒂斯《蓝色裸女》中用钴蓝色来表现丝绸的光滑感[42]。

不同绘画媒介的物理属性，直接制约了色彩的表现方式。宣纸的强吸水性使得水墨能够产生丰富的晕染层次，如赵无极的抽象山水；而亚麻布粗糙的纹理则能承载颜料的厚重笔触，如德·库宁充满力量感的狂暴色块[42]。材料与色彩的结合甚至能产生强大的心理张力，例如，安塞姆·基弗在其《废墟》系列作品中，利用铅皮的冰冷灰调来强化画面的颓败感[42]。

本研究通过建立一套"材质-色彩"的映射规则，揭示特定材质与特定色彩风格或文化语境的关联（如青金石蓝常用于宗教艺术，而荧光色则属于街头涂鸦文化），从而为综合材料艺术的分析拓展出

一套更为丰富的视觉语法。

## 3.6 笔触与肌理生成

笔触与肌理是绘画语言中最具表现力和个性化特征的元素，它们是艺术家身体力行的痕迹，直接承载着情感的流动与精神的印记。本研究将这一复杂的领域系统化，从笔触的功能、形态以及肌理与质感的生成三个层面进行深入剖析，为这些高度个人化的特征提供客观的标注依据。

### 3.6.1 笔触功能

笔触在绘画中扮演着多重角色，它既是技术手段，也是情感的媒介和风格的符号。沈兰[48]将笔触的功能系统地归纳为三个层面，为本知识图谱的分类提供了坚实的理论基础。

首先，笔触是表现客观世界、塑造形体和空间的基本工具[48]。艺术家通过不同形态和走向的笔触来构建物象的结构与动态。例如，梵高用明确而流动的笔触走向来体现《星空》中天体的运动感，而弗洛伊德则用粗阔、堆叠的笔触来描绘人物坚实的肌肉结构。这些笔触服务于"形"的塑造，有力地加强了形的表现意味。

其次，笔触是呈现主观情感与画面节奏的直接载体[48]。不同的笔触能引发截然不同的视觉心理反应。克里姆特作品中跳跃、轻快的笔触传递出画家放达的内心世界，而席勒绘画中那些紧张、滞涩的笔触则让观者窥探到其内心的抑郁与不安。笔触的酣畅或凝重，能让观者直观地体会到心灵的激荡或情感的沉郁，从而烘托出画面的整体氛围与意境。

最后，笔触是形成个人风格符号的关键元素[48]。当笔触的运用超越了单纯的描绘和抒情功能，形成一种稳定且具有高度辨识度的技法语言时，它就成为"画家心灵与个性化的外显"。梵高流动的短线笔触、修拉严谨的点状笔触，都已成为他们艺术风格的独特符号。这些符号化的笔触不仅体现了艺术家对光、色的独特理解，更凸显了其作为一种自觉的艺术创造，在绘画表征上起到了决定性作用。

### 3.6.2 笔触形态

笔触的形态千变万化，其发展演变的历史轨迹反映了绘画观念的深刻变革。本知识图谱将笔触形态系统地划分为书写性笔触、块面笔触、点状笔触和线性笔触四大类。

熊炜[44]在其研究中追溯了油画笔触的发展历史，为这一分类提供了清晰的脉络。从凡·爱克到18世纪的大多数古典画家，普遍采用多层透明画法，有意追求不留笔痕的"工整、细密、光润"效果，这可归为"含蓄笔触"。进入19世纪，德拉克洛瓦开始"以豪放的笔触表达激情"，开启了"奔放笔触"的时代。印象派画家则"用短促破碎的小笔触来描绘物象"，如莫奈笔下颤动的水面，创造了所谓的"颤抖笔触"。梵高在此基础上，发展出连续跳动的粗短笔触和波浪式的旋卷笔触。而新印象派画家修拉和西涅克，则将笔触发展为极具规律性的小色点，形成了独特的"点状笔触"。

此外，许继庄[49]从中西比较的视角指出，西方绘画以团块造型为主（对应"块面笔触"），而中国画则以线造型为主，依靠"书法用笔"产生的粗细、浓淡、疾徐、提按等丰富变化，显示出独特的力感、动感和节奏感。这种具有书法意味的笔触，在本知识图谱中被归纳为"书写性笔触"。这一形态分类体系，不仅涵盖了西方油画史上的主要笔触类型，也融入了中国画的独特笔触美学，具有跨文化的包容性。

### 3.6.3 肌理与质感

肌理是画面表面的物理纹理，而质感则是通过肌理等手段所营造出的可触知的视觉感受。本研究将肌理的生成分为材料综合、物理肌理和视觉肌理三个层面。

物理肌理是直接通过颜料的物理特性和工具的操作形成的。画家常运用颜料的重叠、堆砌、刮擦等手法，使画面产生浮雕般的凹凸感。提香开创的"亮部厚涂，暗部薄染"的技法[50]，极大地增强了光线的效果和画面的生动性。鲁本斯也通过增厚亮部、弱化暗部的方式来增强画面的张力，这种厚涂法成为他塑造对象质感、表现光色的独到手段[52]。

除了涂绘，艺术家还常常利用综合材料来创造独特的肌理，例如在颜料中混合沙子、石膏等，或直接在画布上进行拼贴[50]。

视觉肌理则是通过物理肌理、笔触、色彩等多种手段，在视觉上模拟出特定物体的表面质感。莫奈在《卢昂教堂》系列中，正是通过多次复加颜料的厚涂物理肌理，成功地表现了教堂石墙坚实而粗糙的视觉肌理[44]。伦勃朗对表现肌理尤为偏好，罗杰·弗莱曾评价他"以一种纯真而充满激情的感官之乐探索这些东西"[51]。伦勃朗通过高明暗对比来表现珠宝、盔甲的金属质感，用细致的小笔勾勒来体现斗篷毛茸茸的厚重感，用平滑的笔法处理来表现丝绸的轻薄感[51]。委拉斯贵兹甚至巧妙地利用画布本身的纹理来模拟毛发的质感。这种从物理肌理到视觉肌理的转化，是绘画艺术创造"可触知"幻觉的关键所在。

### 3.6.4 中西方笔触与肌理

中西方笔触与肌理下包含中方画笔触与肌理、西方绘画笔触与肌理、中西方融合笔触与肌理。其中中国画笔触与肌理对应了中国画从"勾勒填色、空勾无皴"到"皴拂阴阳,点匀高下"的皴法再到"水晕墨章"的墨法这一历史演变[59]。

中国画不仅用墨勾勒形体，用皴法塑造山石棱面，更是追求笔墨独特的肌理趣味。西方绘画的笔触与肌理特色主要通过厚薄颜料的对比、捺、点、堆、刷的运笔手型以及麻布帆布木板本身的纹理来展现[44,50]，强调工具对笔触表达与质感塑造的重要性。

19世纪末以来，中西方绘画技法的借鉴融合已成为近现代艺术史中不可忽视的趋势。其一是"取西画写实主义来改造中国，增强描绘现实生活的能力"，比如徐悲鸿画《愚公移山》，用油画的写实技法塑造人物肌肉[60]。其二则是"以西画媒材演绎水墨意境"，比如致力于油画民族化的吴冠中，他的油画作品充满浓郁的江南水乡之诗意，又具现代几何形式和抽象美的风格特点 Error! Reference source not found.。

本研究重视中西方融合对艺术图像发展的多元推动作用，重视现代艺术发展中的多样风格特点，认为此类标签能够帮助艺术图像数据集更具备多样性和前瞻性，提升数据集训练的效果。

### 3.6.5 工具与材料适配

艺术家的个人风格与工具使用、材料特性息息相关。不同画笔的物理形态决定了其笔触语言的范畴，而不同的作画工具是开拓独特视觉肌理的关键。比如格列柯利用硬质粗猪鬃笔能强力剥开已有颜料的特性，达到色彩交融的画面效果；透纳采用调色刀直接将纯色刮在画布表面的作画方式，塑造热烈而富有动感的结构形态[50]。

引入细致的工具与材料标签，能从工具和材料的底层根源解析艺术风格的产生，构建出基于因果链的"工具-材料-效果"可计算艺术创作框架，为多角度数据集标注提供参考。

## 3.7 边缘关系生成

边缘关系的处理是绘画中一项极为精微而关键的技艺，它不仅界定着形状与轮廓，更深刻地影响着画面的空间感、光感、氛围乃至艺术风格的根本属性。本研究将边缘关系作为一个独立的分析维度进行系统化研究，揭示了其在"创造氛围"、"引导焦点"、"界定形状"和"营造空间"等方面的核心功能，并创新性地提出了中西方在边缘处理理念上的根本差异，为这一精微技法的数据化标注提供了可能。

### 3.7.1 边缘清晰度

边缘的清晰度并非一成不变，而是一个从清晰锐利到模糊消失的连续谱系。杨思陶[43]在其论述中指出，边缘线"或柔和、或锐利，或清晰，或模糊"的处理方式，是画家核心技艺的体现，直接服务于艺术表现。本研究将这一谱系系统化为硬边缘、碎边缘、虚边缘和软边缘四大类。

硬边缘，即轮廓明确、界限分明的边缘，常用于古典写实绘画中以塑造坚实的形体，如卢西恩·弗洛伊德"以硬边轮廓线塑造出具有很浓的健美女子意味的女人体"[43]。与之相对，软边缘和虚

边缘则通过模糊、柔和的过渡来表现特定的质感或氛围。例如，雷诺阿"为了表现妇女皮肤柔软温暖的质感，面部轮廓线都被有意画得非常模糊"。印象派绘画更是将这种处理推向极致，通过消除明确的光影区别和轮廓线，来捕捉光线流动、变化的效果[10]。莫奈晚期的《睡莲》系列，以柔韧的笔法打破物象之间的边缘，使景色浑然一体，层次丰富而又通透[44]。碎边缘则介于二者之间，通过不连贯、跳动的笔触来暗示轮廓，常见于印象派和写意风格的作品中。这一清晰度谱系的建立，为量化和标注不同风格的边缘特征提供了精确的词汇。

### 3.7.2 边缘功能

边缘的功能远不止于界定形状。阿恩海姆早已指出，线条（边缘的具象化）不仅能描绘轮廓，甚至可以"再现亮度、空间和空气"[10]。本研究将边缘的功能归纳为创造氛围与光感、引导视觉焦点、界定形状与轮廓、营造空间感四个方面。

在创造氛围与光感方面，画家可以通过虚化边缘来烘托逆光效果[45]，或通过整体的软边缘处理来营造朦胧的雾气氛围。在引导视觉焦点方面，边缘的清晰度对比是有效手段。通常，将主体的边缘处理得比次要物体更清晰（主体硬边缘，次要物软边缘），能够有力地将观者的视线吸引到核心区域。在营造空间感方面，边缘处理遵循着"实前虚后"的基本规律。达·芬奇[27]提出的空气透视理论中就包含了清晰度的梯度变化，即近景清晰、远景模糊。在绘画实践中，将前景物体的边缘处理得相对结实、锐利，而将中远景的边缘处理得虚化、柔和，可以极大地增强画面的前后空间纵深感。

### 3.7.3 中西式边缘处理特色

本研究在边缘关系理论上的核心创新，在于通过对中西方绘画传统的比较，提炼出两种截然不同的边缘处理哲学，并将其概括为中式的"边缘随意境变"与西式的"边缘随光影变"。

西方绘画的边缘处理，自文艺复兴以来，很大程度上遵循着对自然光影的观察与再现。物体的边缘是清晰还是模糊，主要取决于光照条件。例如，伦勃朗、萨金特等人普遍遵循亮部实、暗部虚的边缘处理方法，以戏剧性的光影来塑造鲜明的轮廓。印象派画家虽然模糊了轮廓，但其目的仍是为了更真实地再现光线在物体表面流动、颤动的视觉效果[37,43]。因此，西式边缘处理的根本原则是"随光影变"，其逻辑是感知性的、物理性的。

与此相对，中国画的边缘处理则超越了对客观光影的模仿，完全服务于画面整体的气韵与意境营造。正如黄宾虹所言："古之笔法，今称线条，图画本源，即基于此"[46]。线条（边缘）本身即是独立的审美对象。物体的边缘如何处理，取决于艺术家想要传达的意境。例如，王蒙、倪瓒以清晰的勾线为主，强调山石的结构与笔墨的书写性；而明代画家盛茂烨则用水墨渲染和松散笔触来模糊轮廓，以营造雨水冲击的湿润氛围[47]。因此，中式边缘处理的根本原则是"随意境变"，其逻辑是观念性的、诗性的。

将这两种迥异的边缘处理理念进行提炼和概括，并创造性地提出"中西方融合边缘处理"这一子项，是本知识图谱的重大贡献。它不仅为理解和分析不同文化背景下的艺术作品提供了深刻的理论视角，也为新兴艺术风格（如新中式油画、现代水墨综合材料）的边缘处理模式提供了量化描述的框架，具有重要的理论与应用价值。

## 4 结　论

本研究构建的艺术图像知识图谱，本质是《论视觉知识》[1]提出的"视觉知识结构化"在艺术领域的具象落地，并且补充了当前艺术图像知识图谱在人工智能可读性、可用性、可扩展性上的缺失。

本文系统性地论证了为艺术图像语言与机器学习专门构建的 PL-KG 的理论合理性、科学基础与核心创新价值。通过对构图生成、形状体态生成、透视与空间生成、光影生成、色彩关系生成、笔触与肌理生成及边缘关系生成七大核心维度的逐层解构，本研究表明，该知识图谱并非艺术术语的简单堆砌，而是一套以"中西融合为文化内核、以数据集标注为实践导向、以凸显艺术图像独特性为核心

目标"的严谨、自洽且具备前瞻性的理论框架，为艺术图像的深度分析与标准化标注提供关键支撑。

该知识图谱的核心贡献体现在三个层面。

首先，在文化理论层面，它打破了传统艺术分析中"西方中心"或"东方孤立"的局限，实现了中西方艺术语言的系统性融合与理论重构。该图谱并非简单叠加两种体系，而是从底层逻辑出发，在透视、边缘处理、笔触等多个维度上，找到了中西方艺术的互补性并构建统一的分析框架：在透视与空间维度，将西方基于几何学的线性透视与中国基于哲学意境的"三远法""散点透视"并置，不仅明确二者"固定视点-移步换景""理性空间-情感意境"的本质差异，更创新性提出"融合应用"标签，适配跨文化创作的现代艺术；在边缘处理维度，提炼出"西式边缘随光影变"与"中式边缘随意境变"的核心分野，并增设"中西融合边缘处理"子项，为新中式油画、现代水墨等新兴风格提供标注依据；在笔触与肌理维度，既涵盖西方油画的块面笔触、点状笔触，也纳入中国画"皴法""墨法"的书写性笔触，同时关注新中式油画、现代水墨等新兴跨文化风格，填补了现有艺术分析框架的文化盲区。

其次，在实践应用层面，针对大规模艺术图像数据集标注重缺乏统一标准、主观差异大、难以量化的核心痛点，该知识图谱实现了从定性描述到结构化标注的关键转化，为标注工作提供了可落地、可复用的操作指南。其一，解决定义模糊问题，将传统艺术中抽象的定性概念转化为清晰的层级化标签。例如，将视觉平衡细化为色彩平衡、动态平衡等五类可识别子项，将节奏与韵律拆解为色彩节奏、笔触节律等可观察维度，甚至为留白、散点构图等中式概念明确适用场景，让标注者有统一参照。其二，解决难以量化问题，为核心标签提供可测量的参考指标。例如空气透视对应冷暖变化、清晰度变化、纯度变化三个可量化维度，视觉引导可通过亮度梯度、透视收敛点坐标、黄金分割点位置等辅助判断，使主观美学评价转化为客观可标注特征。这种结构化设计使标注工作摆脱依赖个人经验的困境，鼓励不同标注者在同一话语体系下工作，促进数据标注的一致性与准确性，为后续艺术图像创作智能生成等下游任务奠定高质量数据基础。

最后，在艺术智能层面，它通过针对性的维度设计，精准捕捉艺术图像作为一种特殊视觉媒介的核心特质。与侧重信息记录的一般图像不同，艺术图像的核心价值在于情感表达、风格传递以及观者交互上的独特性。在情感表达层面，图谱未局限于光影、色彩的物理属性，更突出其情感载体功能。如光影维度中，低调绘画对应戏剧性、神秘氛围，高调绘画对应宁静、明亮情绪；色彩维度中，区分东西方共通象征与特色象征，使标注能覆盖"色彩-情感-文化"的完整链路。在风格传递层面，图谱将笔触、肌理等高度个性化的元素设为独立维度，强调其作为"艺术家指纹"的独特性，如梵高的流动短线笔触、修拉的点状笔触、中国画的皴法，这些标签直接关联艺术家风格特征，而一般图像标注极少关注笔触，难以体现艺术创作的主体性。在观者交互层面，图谱聚焦艺术图像的引导性与参与性，如构图中的动线引导、中式留白、散点透视的漫游式体验，这些设计捕捉了艺术图像"主动引导观者解读"的特性，而一般图像仅追求"信息清晰传递"，无交互设计意图。

《文化构成》强调，"文化是生命体，存活在使用中，发展在创新里"，未被现代使用的文化元素如"冰封的生命体"，需通过设计创新使其复活[2]。该艺术图像知识图谱既是对中西方艺术理论的系统性梳理，也是一次"理论服务实践"的创新尝试：其以中西融合突破文化局限，以结构化标签解决标注痛点，以维度设计凸显艺术图像本质，为构建具备"深度艺术语义"的大规模数据集提供了关键工具。未来，可借鉴《论视觉知识》中"全球共建视觉知识词典"[1]的思路，联合艺术史、计算机图形学、AI领域学者，进一步补充非物质文化遗产艺术的视觉标签，完善中西方融合视觉框架，推动艺术视觉知识从数据集标注工具升级为 A 艺术、跨文化研究的基础平台，为 AI 2.0 时代的艺术图像创作智能生成提供关键支撑。

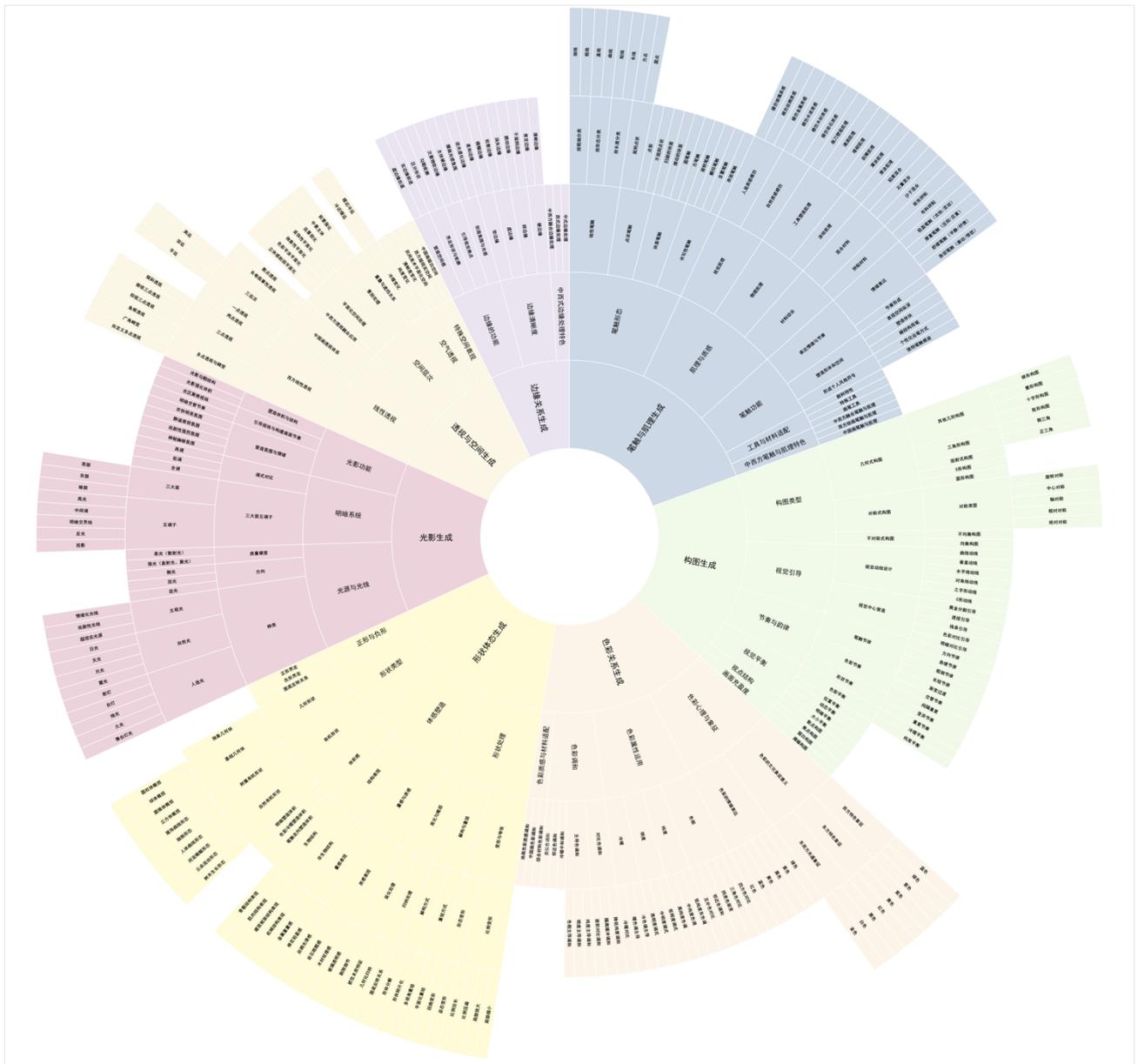

图 1 艺术图像创作标注知识图谱可视化旭日图